\documentstyle[12pt,epsfig]{article}
\makeatletter
\@addtoreset{equation}{section}
\makeatother

\newcommand{\beq}{\begin{equation}}
\newcommand{\eeq}{\end{equation}}
\newcommand{\beqa}{\begin{eqnarray}}
\newcommand{\eeqa}{\end{eqnarray}}
\newcommand{\beqar}{\begin{eqnarray*}}
\newcommand{\eeqar}{\end{eqnarray*}}
\newcommand{\eg}{{\it e.g.,}\ }
\newcommand{\ie}{{\it i.e.,}\ }
\newcommand{\norm}[1]{\raise.3ex\hbox{:}#1\raise.3ex\hbox{:}}

\newcommand{\reef}[1]{(\ref{#1})}
\newcommand{\cR}{{\mathcal R}}
\newcommand{\rb}{r_b}

\begin{document}
\hfill{}
\hfill{}

\hfill{DCPT/01/01, EHU-FT/0012}

\hfill{hep-th/0101198}

\vspace{24pt}

\begin{center}
{\large {\bf Self-gravity of Brane Worlds: A New Hierarchy Twist}}

\vspace{24pt}

Christos Charmousis$^\dag$, Roberto Emparan$^*$\footnote{From Jan
2001, also at
Theory Division, CERN, CH-1211, Geneva 23, Switzerland. E-mail:
roberto.emparan@cern.ch}, and Ruth Gregory$^\sharp$

\vspace{12pt}
{}$^\dag${\sl Institute for Fundamental Theory\\ 
\sl Department of
Physics,  University of Florida}\\ 
{\sl Gainesville FL 32611-8440, USA}\\
{\it charmousis@phys.ufl.edu}\\ 
{}$^*${\sl Departamento de F{\'\i}sica
Te\'orica, Universidad del Pa{\'\i}s Vasco,}\\   
{\sl Apdo.\ 644, E-48080
Bilbao, Spain}\\ 
{\it wtpemgar@lg.ehu.es}\\ 
{}$^\sharp${\sl Centre for
Particle Theory, Durham University, 
South Road, Durham, DH1 3LE, U.K.}\\
{\it R.A.W.Gregory@durham.ac.uk}

\vspace{24pt}

{\bf Abstract}
\end{center}
\begin{quote}{\small

We examine the inclusion of brane self-gravity in brane-world scenarios
with three or more compact extra dimensions. If the brane is a thin,
localized one, then we find that the geometry in its vicinity is warped
in such a way that gravity on the brane can become very weak,
independently of the volume of the extra dimensions. As a consequence,
self-gravity can make the brane structure enter into the determination
of the hierarchy between the Planck scale and a lower fundamental scale.
In an extreme case, one can obtain a novel reformulation of the
hierarchy problem in brane worlds, without the need for large-size extra
dimensions; the hierarchy would be generated when the ratio between the
scales of brane tension and brane thickness is large. In a sense, such a
scenario is half-way between the one of Arkani-Hamed et al.\ (ADD)
(although with TeV-mass Kaluza-Klein states) and that of Randall and
Sundrum (RS1) (but with only a TeV brane, and of positive tension). We
discuss in detail the propagation of fields in the background of this
geometry, and find that no problems appear even if the brane is taken to
be very thin. We also discuss the presence of black branes and black
holes in this setting, and the possibility of having a Planck brane.

}\end{quote}

\medskip

\newpage

\section{Introduction}

The idea that spacetime may have more than four dimensions, with some of them
being macroscopically large, has recently become an attractive scenario where
many a model has been proposed. The common defining features are, on the one
hand, the presence of a three-brane where the Standard Model is confined, and
on the other hand, a bulk where gravity propagates. Other fields may live in
either brane or bulk, as long as their current undetectability is consistent
with experimental bounds. 

The idea dates back at least to \cite{RSh}, but, at the risk of
oversimplification, we may identify two main lines in the recent developments:
one following the works of Arkani-Hamed et al (ADD) \cite{ADD}, and the other
that of Randall and Sundrum (RS) \cite{RS1,RS2}.  Each of these seeks a novel,
geometrical, resolution of the hierarchy problem -- the volume (effective or
real) of the additional dimensions providing the hierarchy between the Planck
and standard model scales. Works under the latter, RS, class typically consider
the brane-world to be a defect of low codimension, say one or two, \ie a
`domain wall' (as in the original RS scenario) or a `cosmic string',
\cite{CK,ruthnsing} and explicitly include the self-gravity (backreaction) of
the brane into the bulk of, typically, an anti-de Sitter spacetime.  By
contrast, the scenario proposed by ADD neglects the self-gravity of the brane,
and regards the number of extra dimensions, $n$, as essentially an adjustable
parameter. In these cases, typically, the extra dimensions have to be
compactified in order to recover the four-dimensional law of gravity at least
down to distances above the experimental bounds \cite{hoyle}. It is quite
possible that in the near future such scenarios with $n<3$ extra dimensions
will be ruled out.

There are in fact good reasons why the bending of spacetime produced by the
brane is dealt with so differently in these two kinds of scenarios. Typically,
a localised defect of codimension one or two manifests its self-gravity in
global features of its spacetime; this is because the intrinsic spatial
directions of the defect do not participate in the gravitational interaction,
and the symmetries of the defect therefore mean that gravity is effectively
propagating only in the directions  orthogonal to it, and of course gravity in
$1+1$ or $2+1$ dimensions does not have any propagating degrees of freedom.
Spacetime exterior to the defect is therefore locally flat, and the
gravitational `field' shows up only globally. For the string, this is the
well-known conical deficit, and for the wall, the spacetime consists of two
portions of flat spacetime (the interior of an accelerating bubble) glued
across their boundaries. Viewed from the frame of the wall itself, the metric
ceases to be static, and inertial observers appear to accelerate away from the
wall. At the location of the `core' of the string or wall, one finds, in the
limit of vanishing thickness, a rather mild distributional  singularity, either
a finite discontinuity in the extrinsic curvature  for the wall, or a simple
apex of the cone spacetime.

Such global, long range effects of low codimension branes are highly
restrictive if one wants to compactify the extra dimensions. This is
unsurprising in the case of the domain wall, since the wall spacetime
itself has compact spatial sections; `standard' compactification with
domain walls requires extra bulk fields \cite{GW}, or walls with
negative tension \cite{RS1}. Compactification with cosmic strings on the
other hand is also problematic, due to the global deficit angle. One
can, in analogy with the wall, introduce a negative tension string for
compactification on a torus, or indeed consider a global, rather than
local, string \cite{CK,ruthnsing} which effectively introduces a bulk
field in the guise of the Goldstone boson of the broken global symmetry.
In addition however, one can consider compactifications on surfaces of
positive total curvature using the deficit angle to provide the
requisite closure \cite{Sun}. However, the fact that no strong
curvatures appear near the sources implies that the thin brane limit
poses no problem, and hence the thickness of the brane can be neglected
to a good approximation. The propagation of fields both in the bulk and
on the brane is very weakly affected by whatever the internal
microstructure of the brane may be. In other words, the low energy
dynamics, in particular the gravity induced on the brane, is quite
independent of any details about the core structure.

Branes of codimension $n=3$, or higher, look in this respect
qualitatively different. If their sources are well localized in the
extra dimensions, then so is the gravitational field they produce: a
localized object of codimension $n$ creates a gravitational potential
that falls off at large transverse distance as $1/r^{n-2}$. Then,
sufficiently far away from the brane, the geometry of spacetime is
hardly affected by the presence of the brane. So, with such branes,
compactification of the extra dimensions presents no special
difficulties, for as long as the thickness of the brane is sufficiently
smaller than the size of the extra dimensions. On the other hand, the
backreaction effects are now shifted to the region close to the brane.
Since the source is localized, in the limit of zero thickness it can be
expected to develop a naked singularity, where the curvature diverges
(horizons surrounding the entire brane are to be avoided). This
singularity will be smoothed out by the physical core of the brane, and
the distance at which this happens depends on the specific details of
the core model. This leaves open the door to the possibility that low
energy physics acquires a strong dependence on the brane thickness, in
contrast to what happened for thin domain walls and cosmic strings. 

For a typical topological defect, such as, say, a cosmic string, the
tension and the thickness are independent quantities, being determined
by different combinations of the parameters of the Lagrangian (the Higgs
self coupling and vev). So there is the possibility that the parameters
that control the brane thickness and brane tension, call them $\rb$ and
$r_0$ respectively, take on very different values. Such an effect,
which is mostly inconsequential for domain walls and strings, will
allow higher codimension branes to generate a hierarchy between the
fundamental scale and the scale of gravity on the brane.
As we will see below, for $\rb$ much
smaller than $r_0$ the relation between the gravitational couplings 
in the bulk and on the brane takes the form
\beq\label{main}
G_4\simeq {G_{4+n}\over V_0}\left({\rb\over r_0}\right)^{{n-2\over
2}\sqrt{n-1\over n+2}}\, ,
\eeq
where $V_0$ is the volume of the extra dimensions, and we assume $r_0$
to be sufficiently smaller (though not necessarily very much smaller)
than $V_0^{1/n}$. 

Equation \reef{main} differs from the standard result from Kaluza-Klein
reduction by the factor involving the brane parameters $r_0$ and $\rb$,
which is absent if the self-gravity of the brane is neglected. This
factor is a consequence of the warping induced by the brane in its
vicinity, and as we see, can have an important effect on the strength of
gravity on the brane. An extreme possibility is that the volume of the
extra dimensions be on the same scale as the fundamental scale (set by
$G_{4+n}$) -- say, around the TeV scale -- and the small value of $G_4$ be
attained by having $\rb/r_0$ small enough.

Hence, even if the starting point resembles the scenario of ADD in that
we have a higher dimensional brane in a compact, empty bulk, the
resolution (or better, reformulation) of the hierarchy problem in the
scenario we have described is perhaps more akin to that proposed in RS1
\cite{RS1}. As in RS1, the hierarchy is generated by the curvature of
the spacetime near the brane, and not by a large internal volume. Also,
the mass of the Kaluza-Klein states is set by $V_0^{-1/n}$. So, if the
latter is on the TeV scale, the large multiplicity of very light
Kaluza-Klein states that is typical of scenarios with large extra
dimensions is avoided.

Note, however, some important differences (besides the different number
of dimensions): in RS1 the hierarchy is generated by an exponential
factor, whereas here we only have a power law. Hence, an ``unnaturally''
large number has to be introduced by hand. Explaining this factor
requires a more fundamental theory (as happens with the large volume
factor in the ADD scenario). Also, in this model there is only one kind
of brane, one of positive tension, which can be regarded as a
``TeV-brane." Gravity becomes Planck-size away from the brane, where the
warping becomes negligible. Finally, the way the warping of spacetime is
generated involves in a crucial way the structure of the brane core, in
particular its thickness. This, as far as we know, is a novel feature in
brane-world scenarios.

Requiring $\rb/r_0$ to be small enough to generate the whole hierarchy
between the TeV and Planck scales is perhaps an extreme situation --
say, for an eleven dimensional universe, $n=7$, a value
$\rb/r_0\sim 10^{-15}$ would be required. In particular, one has to deal
with the very large curvatures induced near the brane. But, at any rate,
the effect encoded in eq.~\reef{main} can have a significant effect in
ADD scenarios, by altering the extra volume required to generate the
hierarchy without changing the energies at which Kaluza-Klein states
start to be excited, the effect being larger for higher values of $n$.
One could envisage, \eg having $\rb$ on the fundamental scale, and $r_0$
being somewhat larger, which would enhance the hierarchy factor for a
given volume. 

The setup we have considered is a minimalistic one, designed to
incorporate the effect of self-gravity of the brane-world into scenarios
like the one of ADD with three or more extra dimensions. To this end, we
consider a brane-world with the following features. First, we shall not
specify any mechanism that gives rise to the brane. The brane will
simply be a thin source (in the limit, a distributional source) of
stress-energy, extended over three spatial directions. It is not clear
whether field-theoretical topological defects can be used to model
this -- to date the only work on
higher codimension brane worlds either uses global defects, \cite{OV},
or is non-specific about the brane core, \cite{GhRS}, indeed there is
some dispute as to whether one can smooth the core without introducing
long range effects \cite{TT}. Nonetheless, the solutions we use as a
model for the self-gravitating higher codimension defect, \cite{CPB}, are 
qualitatively the same even in the presence of Coulomb fields, and so from the 
viewpoint of this paper, we regard this issue as being outside the regime of 
low energy physics. We shall content ourselves if the singularity can be
smoothed, in principle, by a generic brane core. For simplicity, the
only bulk field with non-vanishing background values will be taken to be
the gravitational one, without a bulk cosmological constant. We shall
also require the three-brane to admit a Poincar\'e-invariant ground
state. 

Some of these assumptions might need to be dropped for more fully realistic
model-building, moreover, modification of some of these assumptions may give
rise to qualitative changes. For example, branes in string theory include a
variety of bulk fields which typically do not vanish for the background. In
fact, no branes such as we require are known to exist in string theory, and its
actual p-branes do not share many of the features we have discussed.
Nevertheless, as we will argue near the end, the effect that gives rise to
\reef{main} is rather generic for branes. Hence we think it is important to
study the consequences of this simplest brane-world model. The effects we
describe, and the way we deal with the problems posed by the singularities are
likely to be useful also in other settings.

One final aspect to be mentioned is that, in this framework, gravity is in no
way localized near the brane. This is as in the conventional ADD scenario, and
in order to recover four-dimensional gravity at large distances on the brane,
we shall assume the extra dimensions to be compactified. No specific
compactification scheme will be considered. It might be that a modification of
the model allows for localization of gravity on the brane, \eg by the inclusion
of a cosmological constant, or otherwise. Actually, there have already been
attempts at this. In \cite{GhRS}, a search for self-gravitating, strictly local
defects, such as we consider here, which yield finite gravity on the brane
without compactification, resulted in a negative outcome. Addition of extra
bulk fields appeared to remedy the situation. There are several points of
contact between \cite{GhRS} and our analysis here, but the focus is somewhat
different. Besides this, in \cite{DGP,DG} a model was proposed with a brane in
a flat Minkowski space with an arbitrary number of dimensions. It was claimed
that the model achieved a strong localization of gravity. Self-gravity of the
brane was neglected in that work, but a natural way of incorporating it would
be along the lines described in this paper. More recently, other related recent
work has appeared in \cite{MRK}.

\section{Three-branes of codimension $n$}\label{branen}

Our purpose is to describe three-brane geometries in $4+n$ dimensions.
In the bulk of spacetime only gravity is present, which is described by
the $4+n$ dimensional Einstein-Hilbert action. In addition, there will
be a source of tension $\lambda$ along the four brane-world coordinates
$x^\mu$. This source we take to be localized -- as a first instance, we
take it to be distributional.

Since we require that the brane admits a four-dimensional Poincar\'e
invariant vacuum, for a single brane in a non-compact space we
take the ansatz
\begin{equation}\label{ansatz}
ds^2=A^2(r)\eta_{\mu\nu}dx^\mu dx^\nu+B^2(r)dr^2+C^2(r)r^2d\Omega^2_{n-
1}\, ,
\end{equation}
with $\mu,\nu=0,\dots,3$. 

The cases $n=1,2$ correspond to the well-known domain wall and local cosmic
string geometries. For the domain wall, in an otherwise empty Minkowski space
there are no Poincar\'e invariant umbilic surfaces ($K_{\mu\nu}\propto
g_{\mu\nu}$) with nonzero extrinsic curvature, so walls with the above
symmetries require a bulk negative cosmological constant or additional bulk
fields. For $n=2$, the solution is  analogous to a cosmic string, which creates
a conical deficit angle.  In this case $A=B=C=1$, and the transverse angle is
identified with period $2\pi(1-4 G_6\lambda)$.  

Our interest lies in $n\geq 3$. 
For these cases, the equations have been solved in \cite{CPB},
with the result
\begin{equation}\label{threebrane}
A=f^{-{1\over 4}\sqrt{n-1\over n+2}},\quad B=f^{-{1\over
n-2}\left({n-3\over 2}-\sqrt{n-1\over n+2}\right)},\quad
C=f^{{1\over n-2}\left({1\over 2}+\sqrt{n-1\over n+2}\right)}\, ,
\end{equation}
with
\begin{equation}
f= 1+\left({r_0\over r}\right)^{n-2}\, .
\end{equation}
For example, for the case $n=3$, one gets $A=f^{-1/2\sqrt{10}}$,
$B=f^{2/\sqrt{10}}$, and $C=f^{1/2+2/\sqrt{10}}$.

The parameter $r_0$ is directly related to the brane tension. The relation
between them can be obtained by either reading the energy-momentum tensor of
the brane from the Einstein equations near $r=0$, or by a computation of the
brane tension in an ADM-like approach \cite{robstress}, focusing on the
asymptotic region in the direction transverse to the brane. In this way the
stress tensor is computed to be $T_\mu^\nu = \lambda
\delta_\mu^\nu$, with
\beq
\lambda=\sqrt{(n-1)(n+2)} {\Omega_{n-1}\over 32\pi G_{n+4}}
r_0^{n-2}\, ,
\label{brtension}
\eeq
where $\Omega_{n-1}$ is the area of a unit $(n-1)-$sphere. Since the
space in the directions transverse to the brane is empty, static and
asymptotically flat, this has to be the same as the tension from the
source at $r=0$.

The non-analytic form of the metric coefficients is already an indication
of the presence of singularities. Let us analyze them in more detail.
First of all, the curvature grows to infinity as one approaches $r=0$.
Given the highly localized nature of the source, a naked singularity was
indeed expected. This singularity we identify with the core of the brane.
It was shown in \cite{CPB} that, quite generally, an adequate
(unspecified) choice of core model should smooth out the singularity. We
will assume that smoothing out the core has the effect that the solution
\reef{threebrane} is valid only down to a radius $r=\rb$, and that for
$r<\rb$ the factors $A,B,C$ take values not too different from their
value at $r=\rb$. Hence, the metric induced on the brane is
$A^2(\rb)\eta_{\mu\nu}dx^\mu dx^\nu$.

The space near $r=0$ is highly distorted: notice that $A(r)$ shrinks to zero
at $r=0$. Since $A(r)$ measures the proper size in the directions parallel to
the brane, we see that the brane at $r=0$ has zero size! The singularity is
at a finite proper radial distance, $\int_0 dr B$. In contrast, the size of
the spheres $S^{n-1}$, $rC$, diverge at $r=0$. The latter implies a
significant departure from the situation in flat space, where the transverse
spheres $S^{n-1}$ shrink to zero at the origin. In the present case, it means
that the three-brane is actually delocalized over these spheres. This
delocalization was present also in \cite{GhRS}. Note that the parameter $\rb$
associated to the thickness of the brane is not a proper radius. As a matter
of fact, having a small $\rb$ does not necessarily imply that the brane is
thin: If the thickness is measured by the `area radius', $rC(r)$, then small
$\rb$ means that the brane is actually a thick one! Nevertheless, we will
continue to refer to $\rb$ as the `brane thickness'.

At large distances from the brane the geometry becomes asymptotically
flat, as it typically does for localized objects of codimension $n\geq
3$. Hence, the extra dimensions can be compactified in much the same way
as in the absence of the brane, if the compactification scale $R_c$ is
sufficiently larger than $r_0$ (or $\rb$). In that case, at distances
less than $R_c$ the geometry will be well approximated by
(\ref{threebrane}). To take the compactification effects into account, in
the following we will simply assume that the transverse radial distance
is cutoff at a large distance, $r\leq R_c$. The ratio $r_0/R_c$ may not
need to be too large: one or  two orders of magnitude may be enough.

We end this section by mentioning the relationship between the brane
solutions of \reef{threebrane} and pointlike solutions to
Einstein-scalar gravity in $(n+1)$ dimensions. If we compactify the spatial
brane directions \`a la Kaluza-Klein,
then the remaining spacetime would be of the form
\begin{equation}
ds^2=-f^\alpha dt^2 +f^\beta dr^2+r^2 f^\gamma d\Omega^2_{n-1}
\end{equation}
(in Einstein conformal frame), where,
$$
\alpha=-{1 \over 2}\sqrt{{n+2 \over n-1}},\qquad \beta=-{1 \over n-2}\left(
n-3- {1 \over 2}\sqrt{{n+2 \over n-1}}\right),\qquad \gamma={1 \over n-2}
\left(1+{1 \over 2}\sqrt{{n+2 \over n-1}}\right).
$$
These are static, spherically symmetric solutions to $(n+1)$-dimensional
gravity coupled to a scalar field, the scalar coming from the spatial
volume of the three-brane. For example, for $n=3$ the solutions describe
the singular pointlike objects of \cite{ALC} (see also \cite{wilt} for a
general discussion).

\section{Propagation of scalars}\label{secsing}

The brane thickness and brane tension parameters, $\rb$ and $r_0$, are in
principle independent of each other, and fixed by the core model for the
brane. We will be mostly interested in having $\rb\ll r_0$.

In this case, we have to show that one can consistently define the
propagation of fields in this background for arbitrarily small values of
$\rb$, in particular for $\rb\to 0$. In this respect, the problem is
reminiscent of that of an electron in a Coulomb potential (the hydrogen
atom): even if the source of the potential is a nucleus of finite size
(the analog here of a brane of finite thickness), it is possible to solve
for a singular pointlike source and obtain sensible results by discarding
solutions that are badly behaved at the origin. The Sturm-Liouville
boundary value problem is then well defined\footnote{One needs also to
impose suitable boundary conditions at infinity. We deal with this
problem in a way different from the hydrogen atom, since we consider
space to be compact.}, and unitarity is preserved. After this, the finite
nucleus size can be treated as a small perturbation. Similarly here, the
finite thickness of the brane will result in small corrections to the
propagation of the field in the bulk. 

We shall study the propagation of a massless, minimally coupled scalar
field $\Phi$ in the background of (\ref{threebrane}). Gravitons share
many of the properties of these scalars, but also present important
peculiarities of themselves, and will be the subject of the next section.
Our analysis here will share some technical aspects with the ones in
\cite{CK,GhRS}, although the singularities in those cases lay away from
the core of the brane.

The equation to study is then 
\beq\label{laplacian}
\Box_{4+n}\Phi={1\over\sqrt{-g}}\partial_a\left(\sqrt{-
g}g^{ab}\partial_{b}\Phi\right)=0\, ,
\end{equation}
in the curved background of \reef{threebrane}. 
Given the symmetries of the background, we separate
variables as 
\begin{equation}\label{spheric}
\Phi= e^{ip_\mu x^\mu} Z_{lm_i}(\Omega)\phi_{m l}(r)\, ,
\end{equation}
where $p^2=-m^2$ is the squared mass of the mode from the brane-world
viewpoint, and $Z_{lm_i}(\Omega)$ are the spherical harmonics for the 
$(n-1)$-sphere\footnote{Recall that, even if in higher dimensions there
can be more than one independent angular momenta, in a spherically
symmetric background they are degenerate and the wave  equation depends
only on a single number $l$.}, with eigenvalue $l(l+n-2)$.

The equation \reef{laplacian} is
simplified by noting that $\sqrt{-g}=A^4 B (C
r)^{n-1}\omega_{n-1}=r^{n-1}C^2\omega_{n-1}$
(where $\omega_{n-1}$ is the measure on the
unit $(n-1)$-sphere, which cancels), and $C^2/B^2=f$. The scalar field
equation for
(\ref{spheric})
then reduces to
\begin{equation}\label{laplace}
-{1\over r^{n-1}C^{2}}\left(r^{n-1}f\phi_{ml}'\right)'+{l(l+n-2)\over
r^2C^2}\phi_{ml}={m^2\over A^2}\phi_{ml}\, .
\end{equation}

In order to study the qualitative features of (\ref{laplacian}) 
it is convenient to introduce the `tortoise' coordinate 
\beq
r_*=\int_{0}^r dr{B\over A}\, .
\eeq 
The point $r=0$ corresponds to $r_*=0$, and
asymptotically $r_*\sim r$. We also
change to a new function $\psi_{ml}(r)$ (see \cite{yoabs}),
\begin{equation}
\psi_{ml}(r)=W(r)\phi_{ml}(r),\qquad W(r)\equiv (rC)^{n-1\over 2}
A^{3/2}\, .
\end{equation}
Doing so, we find a Schr{\"o}dinger type of equation for $\psi_{ml}$,
\begin{equation}\label{modeeq}
\left(-{d^2\over dr_*^2}+{1\over W}{d^2W\over dr_*^2}+{A^2\over
r^2C^2}\;l(l+n-2)\right)\psi_{ml}(r)=m^2\psi_{ml}(r)\, .
\end{equation}

\begin{figure}
\centerline{\epsfig{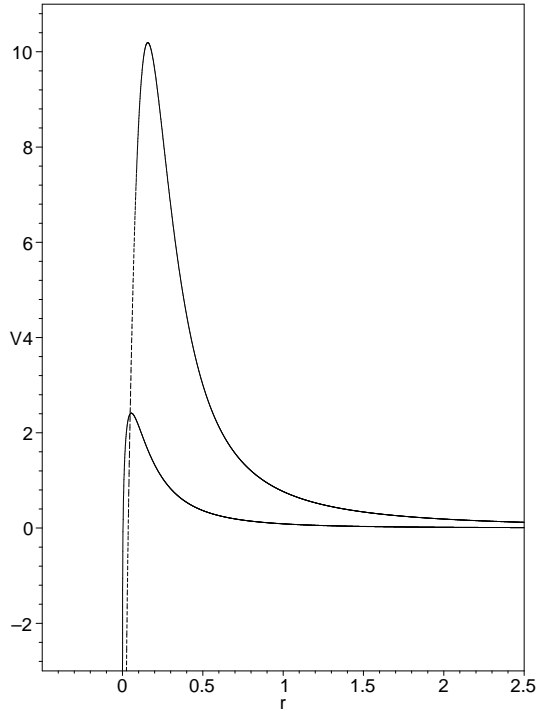}}
\vskip 5mm
\caption{Plot of the potential $V=V(r)$ for $n=3$ (solid line) and $n=4$ 
with $r_0=0.5$}
\label{potential}
\end{figure}

The ``effective potential'' $V$ for the field 
$\psi$ is ${1\over W}{d^2W\over dr_*^2}$, 
which goes to $-\infty$ at $r_*=0$, then reaches a maximum, and
asymptotes to a constant value as $r_*\rightarrow \infty$ 
(see figure (\ref{potential})). 
Near $r_*=0$ the potential diverges, while $A^2\over C^2$ goes to zero.
Hence, all modes will behave near the singularity like the s-wave
component of the massless zero mode, $m=l=0$. Notice that this is
unusual for partial waves with $l>0$, and is in contrast to motion in
flat space, or in the Coulomb potential where, for $l>0$, centrifugal
barriers dominate at short distances. Here, however, the size of the
transverse spheres, $rC$, grows large as $r\to 0$, so the ``angular
velocity'' $\sim J/(rC)$ is small there.

Therefore, in order to investigate the effects of the
singularity we need only to examine the s-wave of the zero mode,
$\phi_{00}$. It
is easy to solve for it as
\begin{equation}
\psi_{00}(r_*)=aW(r_*)+bW(r_*)\log f,
\end{equation}
this is,
\begin{equation}
\phi_{00}(r)=a +b\log f,
\end{equation}
with integration constants $a,b$. Now the question is what boundary
conditions we must impose at $r_*=0$ in order to have a well defined
boundary value problem. Elementary considerations for Sturm-Liouville
problems dictate that
the conditions are that $\phi_{00}(0)$ and $\phi_{00}'(0)$ be
bounded. Hence
we set $b=0$, which implies that the s-wave of the zero-mode is a
constant. The existence
of this solution was clear from the very
beginning, but what is less trivial is that all other modes will behave
similarly near $r=0$. 

To normalize this s-wave zero mode we use the standard Sturm-Liouville
measure in the normalization integral, with the weight function $w(r) =
\omega_{n-1} r^{n-1} C^2/A^2 \ (=\sqrt{-g}/A^2)$ giving the value of $a$
as
\beq\label{zeronorm}
a^{-2}
= \Omega_{n-1}\int dr\; r^{n-1} {C^2\over A^2}\, .
\eeq
This is convergent at its lower limit $r=0$, hence we can get a finite
amplitude even without the need to introduce a cutoff near $r=0$. On the
other hand, the upper bound $r\leq R_c$ is needed in order to render
the amplitude $a$ non-zero. This is not unexpected, since we 
anticipated the need to compactify the extra dimensions.
To leading order in $r_0/R_c$ we have
\beq\label{volume}
a^{-2} \simeq \Omega_{n-1}{R_c^n\over n}\equiv V_0,
\eeq
where, to this order, $V_0$ is the volume of the extra directions. Note
that the actual volume is different, $V=\Omega_{n-1}\int dr\; r^{n-1}
C^2/A^4$. However, both the amplitude of the zero mode on the brane, $a$,
and $V$ are finite even including the singularity at $r=0$, and to
leading order for large $R_c$, $a^{-2}\simeq V\simeq V_0$.

The same boundary conditions can, and must, be imposed for all other
modes. For the massless zero mode, the solution for the higher partial
waves $l>0$, and $n>2$, can be found explicitly as Legendre functions,
$\phi_{0l}\propto  P_{l \over n-2}(x)$ with $x={2r^{n-2} \over 
r_0^{n-2}}+1$ (for $n=3$ we get Legendre  polynomials $P_l$). These modes
take on non-zero values at $r=0$, but they grow as $r^l$ at large $r$, as
in the absence of brane self-gravity. They will be eliminated once the
boundary conditions for compactification are imposed.

\begin{figure}[t]
\centerline{\epsfig{file=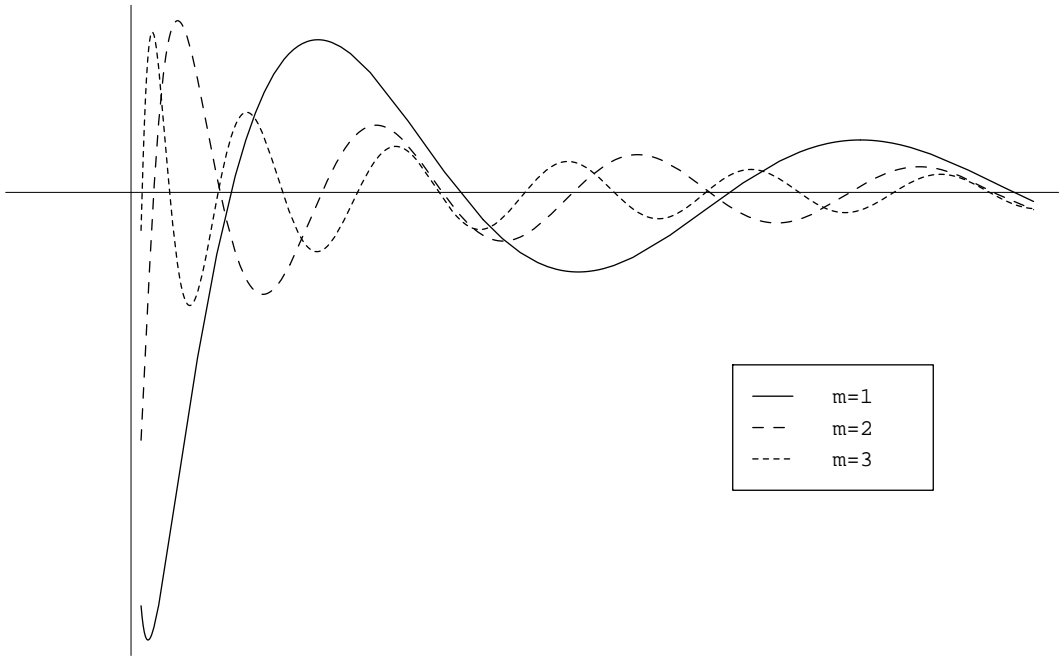,width=14cm}}
\caption{Plot of massive mode wavefunctions for $n=3$, $r_0=1$.}
\label{harmonics}
\end{figure}

Massive modes also tend to a finite value at $r=0$. At larger values of $r$
their radial wavefunctions asymptote to their flat space form in terms of
Bessel functions, $\sim r^{1-n/2}J_{n+2l-2\over 2}(mr)$. Again, only the s-wave
component is expected to remain after compactification. The shape of some of
these modes is exhibited in figure (\ref{harmonics}).

With these results in hand, we are ready to study the interactions mediated by
scalars, on either brane or bulk. The (Euclidean)
scalar two-point Green's function between points with brane and bulk
coordinates $(x^\mu,{\bf r})$ and $(x^{\mu\prime},{\bf r}')$, 
respectively, can be computed from the (normalized) eigenmodes $\phi_{ml}$ as
\beq\label{propag}
G(x,x';{\bf r},{\bf r}')=\int {d^4p\over
(2\pi)^4}e^{ip_\mu(x^\mu-{x^\mu}')}\sum_{m=0}^\infty
\sum_{l,m_i}Z_{lm_i}(\Omega)Z_{lm_i}(\Omega')
{\phi_{ml}(r)\phi_{ml}(r')\over p^2+m^2}\, .
\eeq
Compactification will eliminate higher partial waves, and will also make
the mass spectrum discrete. Then, for two points on the brane at a large
distance, the effective interaction potential will be predominantly
mediated by the zero mode, and be given by 
\beq\label{pot}
{\phi_{00}(0)^2\over |x-x'|}\simeq {1\over V_0}{1\over |x-x'|}\, .
\eeq
Corrections to the leading order value $\phi_{00}^2\simeq V_0^{-1}$ will
appear at order $(r_0/R_c)^{n-2}$. The discrete massive modes will, as usual,
give finite corrections of Yukawa type.

Finally, we can also study the stability of the perturbations. The
argument is a
standard one. First, write \reef{modeeq} as 
\beq
\left({d\over dr_*}+{d\log W\over dr_*}\right)
\left(-{d\over
dr_*}+{d\log W\over dr_*}\right)\psi_{ml}(r)+{A^2\over
r^2C^2}l(l+n-2)\psi_{ml}(r)=m^2\psi_{ml}(r)\, .
\eeq
With the boundary conditions specified above, the operators
${d\over dr_*}+{d\log W\over dr_*}$ and $-{d\over
dr_*}+{d\log W\over dr_*}$ are each the adjoint of the other. Hence the
operator that acts on $\psi_{ml}$ on the left hand side of (\ref{modeeq}) is
a positive semidefinite self-adjoint operator. Its eigenmodes then
provide a complete basis, and their eigenvalues $m^2$, are all non-negative.
These perturbations are, then, stable. 

Hence we have proven that even if the background metric exhibits a naked
singularity at the position of the brane, the propagation of minimal scalars,
and the interactions mediated by them, on either brane or bulk, are well
defined. The inclusion of a small but finite brane thickness cutoff $\rb$ can
then be treated as a small perturbation.

In this sense, we have shown that the solutions \reef{ansatz},
\reef{threebrane} are the genuine cousins of the conical string, and
distributional wall, spacetimes --  in that the nominal singularity of
the metric is quite integrable as far as causal propagation of fields on
the spacetime is concerned.

\section{Gravity on the brane}

The leading static scalar potential \reef{pot} appears to be finite and largely
independent of the cutoff imposed by the thickness of the brane\footnote{Our
views and interpretation in this section have been greatly improved by comments
from Valery Rubakov.}. However, notice that, in \reef{pot}, the distance
$|x-x'|$ is a coordinate distance,
whereas the proper distance would be $A(\rb)|x-x'|$. This gets smaller as $\rb$
decreases. Assuming, as we said, that on the brane the functions $A,B,C$ do not
vary appreciably from their value at $r=\rb$, then the
proper distance between two points on the brane will be $\ell=A(\rb)|x-x'|$,
and the
leading order static interaction potential
\beq
V(x,x')={g_{4+n} A(\rb)\over V_0}{1\over \ell}\, ,
\eeq
where $g_{4+n}$ is the scalar coupling in $4+n$ dimensions. Hence it appears
that the effective coupling on the brane, $g_4$, takes the form
\beq\label{couplings}
g_4={g_{4+n} A(\rb)\over V_0}\, .
\eeq
This is different from the standard result in the absence of brane
self-gravity, which would not have the factor $A(\rb)$. As $\rb$ gets
smaller, the coupling on the brane gets weaker, and vanishes for a brane of
zero thickness. Therefore, self-gravity makes the effective coupling induced
on the brane acquire a strong dependence on the brane thickness.

Indeed, this feature would have been expected not only for scalars, but also
for the gravity that is induced on the brane. Typically, in a brane world
model, gravity becomes weaker where the brane metric is smaller, in the sense
that the conformal factor $A^2$ in front of the metric becomes
smaller\footnote{This is also familiar from string theory, where the
supergravity solutions for fundamental strings or D-branes show that they can
be coupled perturbatively (weakly) to gravity, \ie to closed strings.}. In
the present case, the proper size of the brane is proportional to $A(\rb)$,
which decreases as $\rb$ gets smaller. 

Let us see in more detail how this comes about. To study the propagation
of gravitons in the background of these branes, we perturb the metric
$g_{ab}=g_{ab}^0+h_{ab}$, where $g^0$ is the background metric of
\reef{threebrane}, and we choose a gauge where perturbations are transverse and
trace-free (TTF). The perturbations with indices along the brane metric,
$h_{\mu\nu}$, will be related to tensor gravity on the brane. The brane scalars
$h_{rr},h_{\theta\theta},h_{r\theta}$ give rise to massless scalars, which,
upon compactification, correspond to the moduli for deformations of the
internal space. These are a source of phenomenological difficulties, common to
any compactification scheme. In the following, (in keeping with much of the
literature) we shall assume that there exists a mechanism that gives a large
mass to the moduli and stabilizes them up to some high energy scale. If this is
the case, they will decouple at low energies, so these components of $h_{ab}$
can be set to zero. We will come back to this point in the final
section.

With these assumptions, it is enough to perturb
$A^2\eta_{\mu\nu}\rightarrow A^2\eta_{\mu\nu}+h_{\mu\nu}$
(in TTF gauge, $\nabla_\mu h^{\mu\nu}=h_\mu^\mu=0$). Let us factor out the
conformal factor and define $\hat
h_{\mu\nu}=A^{-2}(r)h_{\mu\nu}$. The
resulting Lichnerowicz equation is
\begin{equation}
\left[ \Box_x \hat h_{\mu\nu} + 2 R_{_0\mu\lambda\nu\sigma}\hat
h^{\lambda\sigma} \right] - {A^2\over\sqrt{-g}}
\left[ {\sqrt{-g}\over B^2} \hat h_{\mu\nu}' \right]'=0\, ,
\end{equation}
where primes denote $r$-derivatives. The polarization structure can also
be factored out, $\hat h_{\mu\nu}(x,z)=\epsilon_{\mu\nu}\Phi(x,z)$, with
$\epsilon_{\mu\nu}$ a constant polarization tensor. Doing this, the equation
that results for $\Phi$ is precisely the same as for a minimally coupled
scalar, which we
have studied in the previous section. Hence we find that, in a compact space,
the interactions are dominated by the s-wave of the zero mode of the graviton.
In order to determine the gravitational coupling on the brane it is then enough
to focus on the latter mode.

In fact, the easiest way to do so is by working with the s-wave of the zero
mode of the graviton exactly, \ie beyond the linearized approximation. This is
easy, since that mode is constant. Non-perturbatively, this is to say that the
metric (\ref{threebrane}) continues to be a solution (with the same functions
$A,B,C$, and $f$), when instead of the Minkowski metric we place an arbitrary
Ricci-flat metric $\hat\gamma_{\mu\nu}(x)$,
\begin{equation}\label{zeromode}
ds^2=A^2(r)\hat\gamma_{\mu\nu}(x)dx^\mu
dx^\nu+B^2(r)dr^2+C^2(r)r^2d\Omega^2_{n-1}\, .
\end{equation}
This $\hat\gamma_{\mu\nu}(x)$ is the non-perturbative counterpart of the
s-wave of the zero
mode of the perturbation $\hat h_{\mu\nu}$. Actually, the metric induced on the
brane is not $\hat\gamma_{\mu\nu}(x)$, but instead
$\gamma_{\mu\nu}(x)=A^2(\rb)\hat\gamma_{\mu\nu}(x)$, hence we write
\begin{equation}\label{zeromode2}
ds^2={A^2(r)\over A^2(\rb)}\gamma_{\mu\nu}(x)dx^\mu
dx^\nu+B^2(r)dr^2+C^2(r)r^2d\Omega^2_{n-1}\, .
\end{equation}

The $(\mu,\nu)$ components of the Ricci tensor in the
bulk $R_{\mu\nu}$, are equal to those of the Ricci tensor in the brane,
$\cR_{\mu\nu}(\gamma)$,
so
\begin{equation}\label{scurvs}
R={A^2(\rb)\over A^2(r)}\cR\, .
\end{equation}
The relationship between the
Newton constant in the bulk and in the brane follows easily now by
integrating the extra dimensions in the Einstein-Hilbert action,
\beq
{1\over G_{4+n}}\int dr\; d\Omega^{n-1}\; d^4x \sqrt{-g}R={1\over G_{4}}\int
d^4x \sqrt{-\gamma}\cR +\dots, 
\eeq
using \reef{scurvs},
\beqa\label{ehterms}
{G_{4+n}\over G_4}&=&\int
d\Omega^{n-1} dr\; {A^2(\rb)\sqrt{-g}\over
A^2(r)\sqrt{-\gamma}}={1\over A(\rb)^2}\int
d\Omega^{n-1} dr\; r^{n-1} \; {C^2\over
A^2}\nonumber\\
&\simeq&{V_0\over A^2(\rb)}\, ,
\eeqa
where, up to the constant $A^2(\rb)$, the integral is the same as
appeared in eqs.~\reef{zeronorm} and
\reef{volume}, and is evaluated under the same approximations. For $\rb\ll r_0$
and using the expression for $A$ in \reef{threebrane}, we obtain the result we
announced in the introduction \reef{main},
\beq
G_4\simeq {G_{4+n}\over V_0}\left({\rb\over r_0}\right)^{{n-2\over
2}\sqrt{n-1\over n+2}}\, .
\eeq
This is our main result. It implies that the dimensionless ratio
$\rb/r_0$ can be used for, or contribute to, generating a hierarchy
between the fundamental scale of $G_{4+n}$ and the Planck scale on the
brane of $G_4$. There are several possibilities. One is that the whole
hierarchy is generated by this ratio, with $R_c(\sim V_0^{1/n})$ being
at the fundamental scale. In this case, the hierarchy problem consists
of explaining why the brane thickness and tension have such an
unnaturally small ratio\footnote{{\it I.e.}, small $\rb/r_0$, but we
remind the reader that in a certain sense small $\rb$ implies a thick
brane}. Another possibility is that the hierarchy is partly generated by
the volume factor, and partly by the brane parameters ratio. Since the
curvature near the brane is controlled by $\rb$, the latter might be at
the fundamental scale, while $r_0$ might be somewhat different. The
interesting feature of this result is that the hierarchy can be changed
without significantly varying the mass of the Kaluza-Klein states, which
is determined by $1/R_c$.

Finally, it might perhaps be argued that the factor
$(\rb/r_0)^{{n-2\over 2}\sqrt{n-1\over n+2}}$ could be absorbed by a
redefinition of the internal volume. However, we have seen that the
actual internal volume is finite even for $\rb=0$, and moreover, that
would not be a useful viewpoint, since the mass of the Kaluza-Klein
states is set by $V_0^{-1/n}$. The extra factor comes essentially from
the curvature scalar $R$. Again, this is similar to the situation in the
RS1 model.

\section{Black branes and black holes on the brane}

By exciting only the s-wave of the zero mode one can construct {\it
black $n$-branes}. If the effect of compactification is neglected, they
correspond to
\begin{equation}
ds^2=A^2(r)\left[-\left(1-{2m\over \rho}\right)dt^2+{d\rho^2\over
1-{2m\over \rho}}+\rho^2d\Omega^2_2\right]+B^2(r)dr^2+C^2(r)r^2d\Omega^2_{n-1}
\label{blackstr}
\end{equation}
(taking, again, (\ref{threebrane}) for
$A,B,C$)\footnote{As a side remark, note that this solution describes a
localized intersection of branes, \ie the three-brane (the brane-world)
is intersected by the black $n$-brane that extends in the directions
transverse to the former. It is expected by no-hair arguments that such
intersections are singular, as is so in this case.}. The size of the
horizon in the directions transverse to the brane grows as $r=0$ is
approached, while in the parallel directions it shrinks. The two effects
considered, the overall size grows, but the
entropy remains finite: the area of the horizon of the black brane is
\beq
{\cal A}=\Omega_{n-1} \int dr\; r^{n-1}A^3 C^{n-1}=\Omega_{n-1} \int
dr\; r^{n-1}{C^2\over AB}\, ,
\eeq
and it is easy to see that even if the integrand diverges at $r=0$,
the divergence can be integrated.
Hence, the entropy of these black branes is finite (if in a compact
space) and it gets somewhat higher near the brane-world than it would be
if the bulk were not curved by the brane-world.

This solution, like
its counterparts in the absence of the 3-brane \cite{GL}, or in anti-de
Sitter space \cite{G} is in fact unstable if the size of the horizon, 
$2 m$, gets smaller than the compactification length. Recall that in \cite{GL},
it was shown that a black string (metric (\ref{blackstr}) with $A=B=C=1$)
is unstable to fragmentation in the extra ${\bf r}_n$ dimensions.
The instability takes the form of a 4D massive tensorial TTF s-wave
perturbation:
\beq
h_{ab} = e^{i\mu\cdot r} \left [ \matrix{ \matrix{ 
{\tilde h}_{tt} & {\tilde h}_{t\rho} &0&0\cr
{\tilde h}_{t\rho} & {\tilde h}_{\rho\rho} &0&0\cr
0&0&{\tilde h}_{\theta\theta} & 0 \cr
0&0&0 & {\tilde h}_{\theta\theta} \sin^2\theta \cr} 
& {\bf 0} \cr ~ & ~ \cr ~ & ~ \cr
{\bf 0} & {\bf 0}\cr}\right ]\, .
\eeq 
The effect of this instability (due to the $e^{i\mu\cdot r}$) is to induce a 
rippling in the event horizon -- which by entropic arguments we expect to 
fragment. The key feature of this instability (demonstrated and exploited in
\cite{G}) is that it lies purely within the 4D $\mu-\nu$ coordinates,
and the action of the full Lichnerowicz operator produces a 4D Lichnerowicz
operator with a mass term arising from the prefactor depending on the 
orthogonal directions
\beq
\Delta_{n+4} h_{ab} \propto \left (\Delta_4  + 
M^2 \right ){\tilde h}_{\mu\nu} \, .
\label{lichsplit}
\eeq
In the case of flat extra dimensions, this observation is trivial, however,
if the extra dimensions are curved, as in (\ref{blackstr}) (or as in
the RS scenario \cite{G}) this splitting in fact still occurs.
Therefore, the general instability will have the form
$f_{_{\rm M}}({\bf x}_\perp) {\tilde h}_{\mu\nu}$, where ${\tilde h}_{\mu\nu}$
is the massive tensorial instability of (\ref{lichsplit}) and
$f_{_{\rm M}}({\bf x}_\perp)$, is the appropriate orthogonal eigenfunction
giving rise to that mass. So, for the
standard p-brane instability of \cite{GL}, where the transverse
dimensions are flat, the eigenfunction is $e^{i\mu_i x^i}$; for
anti-de Sitter space, it is the appropriate Randall-Sundrum
orthogonal eigenfunction $u_m(z)$ \cite{RS2,G}. In our case, we have
demonstrated the eigenfunctions $\phi_{ml}$ in section \ref{secsing},
and therefore the appropriate form of the instability is
\begin{equation}
h_{\mu\nu} \sim A^2(r) Z_{lm_i} (\Omega) \phi_{ml}(r) 
{\tilde h}_{\mu\nu}\, .
\end{equation}
The horizon therefore develops a instability, rippling with a
shape given by the waveforms in figure \ref{harmonics}.

The black brane in a large enough compact space is expected to collapse
into a black hole. How to get an exact description of the black hole on
the brane remains an open problem for $n\geq 3$, even when  brane
self-gravity is neglected\footnote{Toy models for lower  dimensional
branes are more tractable, see \eg \cite{EHM1}.}. In the latter
approximation, and if the finite compactification size effects can be
neglected, then the black hole is well approximated by the $4+n$
dimensional Schwarzschild solution, at least for distances on the brane
that are smaller than the compactification length. In coordinates adapted
to the presence of the brane, this is
\beq
ds^2=\eta_{\mu\nu}dx^\mu dx^\nu +dr^2+r^2d\Omega_{n-1}^2+{2m\over
(r^2+x^i x^i)^{(n-2)/2}}dt^2-
{2m\over (r^2+x^i x^i)^{n/2}}(rdr+x^i dx^i)^2
\eeq
($i=1,2,3$). The brane corresponds to the section $r=0$.

Brane self-gravity will significantly distort this geometry near the core of
the brane. In particular, we have seen that the size of the spheres $S^{n-1}$
grows larger near the brane, and the directions parallel to the brane to
shrink. Hence, we can expect that the black hole will pinch at the brane core
in directions parallel to the brane, but will greatly expand in the directions
transverse to it. As was the case for the black brane, these two competing
effects are expected to balance in the calculation of the area of the horizon
in the bulk, which is expected to remain finite. In general, as argued in
\cite{EHM3}, black holes remain attached to the brane. In the present case this
may be even more so, since the area of the horizons gets comparatively higher
near the brane. 

\section{Discussion}

Let us now discuss, mostly from a qualitative point of view, an issue we
have left aside so far, namely the compactification moduli. Even before
dealing with their stabilization, we should expect the moduli amplitudes
to be very large at the position of the brane. It was already observed in
\cite{CGR} that in the RS1 model, the corresponding modulus (the radion)
gets large there where gravitons parallel to the brane get small, and
vice versa. This is, in fact, a consequence of Einstein's equations in
vacuo. In our case, it would be in accord with the fact that the metric
in the directions transverse to the brane grows without bound as $\rb$
gets smaller, while it goes to zero in the parallel directions.

Now suppose that we compactify the extra dimensions and somehow manage to
stabilize the moduli and give them a large mass. The infinite (for
$r_b=0$) amplitude of the moduli at the brane would seem to imply that
the brane is extremely sensitive to (scalar) deformations of the compact
manifold. In order to avoid this, the compact manifold has to be very
rigid, \ie the mass of the moduli has to be very large. With finite brane
thickness, the minimum mass of the moduli can be expected to be
determined by, roughly, $1/\rb$. Since the equations for parallel
gravitons and moduli do not decouple, any fluctuations in the moduli will
have an effect on the parallel gravitons. Therefore, the question of
moduli stabilization is actually of relevance to the entire setup.
Nevertheless, we have seen that if it can be appropriately solved, then
all other issues about the consequences of brane self-gravity and the
influence of the brane thickness can be safely dealt with.

Although we have been considering the specific case of what is arguably the
simplest three-brane solution, the weakening of gravity on the brane is
actually rather generic. Recall that the warp that gives rise to the factor
$r_b/r_0$ in \reef{main} is due to the shrinkage of the metric along the
directions parallel to the brane, while the brane core remains at a finite
proper spatial distance. This shrinkage is, in fact, a generic consequence of
the tension along the brane directions. Hence, once one has a brane with a
certain positive tension, the same shrinkage will happen even in the presence
of extra bulk fields, for as long as the gravity of the latter is not so
strong as to upset the effect. In fact, the non-generic cases will be those
for which the extra bulk fields act in such a way as to exactly
counterbalance the shrinkage of the geometry -- such is the case of \eg
D3-branes, for which the gravity of the RR field has the effect of pushing
the brane core down an infinite throat.

To conclude, we have shown how to incorporate the gravitational field
induced by a brane with three or more transverse dimensions, and have
found that it can lead to important effects. Although the solutions
\reef{ansatz}, \reef{threebrane} exhibit singularities at the core of the
brane, the singularities are such that bulk fields can propagate in their
presence in a finite and unitary way. Hence, we can legitimately regard
these solutions as the higher dimensional analogues of the domain wall
and cosmic string spacetimes. Concerning the gravity that is induced on
the brane, a new dimensionless parameter, which is determined by the
brane structure, enters into the determination of the effective four
dimensional gravitational constant. This can be used to generate, or
contribute to, a hierarchy between the Planck scale and a fundamental
scale. Nevertheless, the energy scale for Kaluza-Klein excitations
remains largely unaffected by variations of that parameter. Such an
influence of brane parameters on the hierarchy is a novel effect, to be
added to ADD scenarios. It mixes in the main ingredient present in the
RS1 model, namely, the warping created by the brane self-gravity, and
hence provides the possibility of taking advantage of some of the
features of each model. 

Of course, if one wishes instead to deal with a genuinely fat brane-world,
one must first find a reasonable core model. Ghergetta et.\ al.\ \cite{GhRS}
point out some problems with a fully localised core, and it is quite possible
that none such exists. If one wishes to model the brane by a topological
defect (as in \cite{OV}) then one generally has a spontaneously broken
symmetry in the vacuum of the theory. For a defect of codimension $n$, this
mandates a nontrivial $\Pi_{n-1}$ homotopy group of the vacuum  manifold (or
coset space $G/H$). For simple Lie groups, this is equivalent to
$\Pi_{n-2}(H)$, the unbroken group, but if $H$ is nontrivial, then there will
be Coulomb fields surrounding the core of the defect \cite{TT} -- in other
words, we might expect any smooth brane-world {\it not} to have a fully
localised core.  
Another possibility is that the brane is not a
field-theoretical defect, but a brane originating from string theory. It
cannot be a D3-brane, but the latter is in fact a very special case due to
conformal invariance. More realistic stringy brane-worlds may well exhibit
the kind of warping we have described. 

Indeed, string theory suggests a way in which a small ratio between brane
thickness and brane tension can arise. Suppose that string theory, or
whatever the fundamental theory is, has branes of the form we require (or
similar, in that they present the appropriate shrinkage in brane directions
while keeping the brane core at a finite proper distance). For a single brane
in that theory the tension and thickness scales would both be at the
fundamental scale, say the string scale $l_s$. Now, if instead of one brane
we have a stack of $N$ branes, then the thickness would presumably still be
$r_b\sim \l_s$, but since we have $N$ branes the tension would have grown. In
the absence of interactions between the branes one would have, if
eq.~\reef{brtension} holds, $r_0\sim N^{1/(n-2)} l_s$. Interactions betwen
the branes, if attractive, will reduce this value, but at any rate the
tension will grow as $N$ grows. Therefore $r_0/r_b$ grows with
$N$, and a large hierarchy appears for large $N$. A number of issues has to
be dealt with, since having a stack with a large number of branes increases
accordingly the number of brane fields, including the light ones. A
way to prevent them from interfering with the fields we actually observe
would have to be provided. The question of whether this is a workable
direction remains an open one.

For now, we cannot fully answer all these questions in detail, but we simply
note that we have shown that the brane-world gravity can be dramatically
dependent on the specific model used to generate the core.

\section*{Acknowledgements}

We would like to thank Vakif Onemli, Pierre Ramond, Valery Rubakov,
Tigran Tchrakian and Richard Woodard 
for helpful discussions.

CC is supported by US DoE grant DE-FG02-97ER41209,
RE is partially supported by UPV grant 063.310-EB187/98 and
CICYT AEN99-0315, and RG is by the Royal Society. We would
also like to acknowledge the support of a Royal Society ESEP
exchange grant.

\appendix

\section{How to obtain a ``Planck brane''}

If the effect of brane self-gravity is used to generate the entire
hierarchy between the TeV and Planck scales, then, in the terminology of
\cite{RS1} the brane would be a ``TeV-brane.'' In contrast to \cite{RS1},
there is no ``Planck-brane'' in this model. A ``probe'' Planck brane
would have to sit at a location where the metric factor $A(r)$ takes on
values of order unity. 

Nevertheless, there is a way to make the self-gravitating brane of
\reef{threebrane} a Planck brane. Indeed, it allows to find finite gravity on
the brane even in the limit $\rb\to 0$, as occurs for the domain wall or
cosmic string brane-worlds. In that case, the limit where the brane has zero
size is sensible on its own, and the physics at low energies is actually
independent of any details of the brane structure. This might be a desirable
feature.

The procedure to achieve this lies on an alternative, if somewhat {\it
ad hoc} definition of what the brane metric is, \ie to which metric does
matter on the brane couple to. Usually, fields that live on the brane
are just a truncation of fields that actually can extend into the bulk,
but whose propagation into the bulk is heavily suppressed, by say, the
topology of the field configuration. This is, they are actually bulk
fields, and hence they will couple to gravity through the metric induced
on the brane. Now, let us suppose instead that the fields that live on
the brane do not couple to the metric induced on the brane (which has
zero size if $\rb=0$), but rather to one that is related by a constant
rescaling. This implies rescaling the strength of the gravitational
interactions on the brane, which can be done in such a way that a finite
brane metric results, and finite gravity on it, even if $\rb\to 0$. More
specifically, in the spacetime \reef{zeromode}
\begin{equation}
ds^2=A^2(r)\hat\gamma_{\mu\nu}(x,r,\Omega)dx^\mu
dx^\nu+B^2(r)dr^2+C^2(r)r^2d\Omega^2_{n-1}\, ,
\end{equation}
let us {\it define} our brane metric to be
$\hat\gamma_{\mu\nu}(x)\equiv\hat\gamma_{\mu\nu}(x,r=0)$. By this we mean that
matter fields on the brane will couple to
gravity through the metric $\hat\gamma_{\mu\nu}$. For example, we take the
action for a brane-scalar $\varphi$ as
\begin{equation}
{1\over 2}\int d^4x \sqrt{-\hat\gamma}(\hat\gamma^{\mu\nu}\partial_\mu
\varphi\partial_\nu \varphi+\mu^2\varphi^2)\, .
\end{equation}
Equivalently, in the linearized approximation where
$\hat\gamma_{\mu\nu}=\eta_{\mu\nu}+\hat h_{\mu\nu}$, the coupling to a
brane source of stress-energy $T_{\mu\nu}$ will be through a term
\begin{equation}\label{branecoup}
{1\over 2}\int d^4 x T_{\mu\nu} \hat h^{\mu\nu}\, .
\end{equation}
This is different to coupling to the  actual metric perturbation
$h_{\mu\nu}=A^2(r)\hat h_{\mu\nu}$. In fact, a coupling of the form
\reef{branecoup} does not arise naturally in any of the common ways to
generate brane fields from a topological defect. However, one may argue
that the question of how to generate such couplings belongs in the realm
of the higher theory that gives rise to the brane, and which we have
explicitly left out of our discussion.

It is now a straightforward matter to show, following the steps that lead to
\reef{main}, that to leading order in $r_0/R_c$, the
Newton constant on the brane is given by the standard formula,
\beq\label{gfour}
G_4\simeq {G_{4+n}\over V_0}\, ,
\eeq
instead of \reef{main}. Notice that even if $\hat h_{\mu\nu}$ satisfies
the same equation as the scalar $\Phi$ of the previous section, the
argument that the distance $|x-x'|$ that appeared in the interaction
potential \reef{pot} is not the proper distance is not relevant anymore.
For, if brane matter couples to the metric $\hat\gamma_{\mu\nu}$, then 
$|x-x'|$ is the distance actually measured by a brane observer (with
rulers made of brane matter), and the interaction potential is indeed of
the form \reef{pot}, without any extra factors.

We can now analyze the different gravitational interactions between brane
and bulk matter. We have seen that $\hat h_{\mu\nu}$ satisfies the same
equation as the minimal scalar $\Phi$ of the previous section. Hence, for
two sources on the brane, which couple to gravity through the interaction
term \reef{branecoup}, the relevant two-point function is
\beq
\langle \hat h_{\mu\nu}(x)\hat
h_{\rho\sigma}(x')\rangle=\epsilon_{\mu\nu}\epsilon'_{\rho\sigma}
G(x,x';0,0) \, ,
\eeq
where $G(x,x';0,0)$ is that of \reef{propag} with the two points at
$r=0$. This is finite, and all the comments made about the scalar
interaction on the brane through \reef{propag}, apply here as well. 

Now, if one of the sources is on the brane and the other is in the bulk,
with the stress-energy tensor of the latter lying along the brane
directions, and coupling as $\sim\int T_{\mu\nu}h^{\mu\nu}\sim \int
A^2(r)T_{\mu\nu}\hat h^{\mu\nu}$, then
\begin{equation}
\langle h_{\mu\nu}(x,z) \hat h_{\rho\sigma}(x',0)\rangle=\epsilon_{\mu\nu}
\epsilon'_{\rho\sigma}A^2(r)G(x,x';z,0)\, .
\end{equation}
Again, this is finite, although it becomes smaller as the bulk source
gets closer to the brane. 

This serves us to illustrate one of the important features of this
prescription for obtaining a Planck brane, namely, that brane matter and
bulk matter are, with this prescription, fundamentally different. The
gravitational interactions of matter in the bulk (at least those
mediated by gravitons with polarization parallel to the brane) become
weaker closer to the brane. At $r=0$, its coupling to gravity vanishes.
In contrast, brane matter gravitates finitely, but if one tried to pull
it out of the brane its weight would presumably become infinite. Hence
such brane matter would have to be permanently confined to the brane.
Bizarre as this may sound, the prescription appears to be a consistent
one, for which the issue of the correct coupling of gravity can only be
resolved from the point of view of the theory that underlies the brane
structure. The most striking consequence of this is that finite results
can be obtained even if the proper size of the brane has shrunk to zero!

\end{document}